\documentstyle[amsfonts,tighten,preprint,eqsecnum,aps,prd]{revtex}
\begin{document}
\draft

\baselineskip=16pt  

\hyphenation{
mani-fold
mani-folds
}


\def\BbbR{{\Bbb R}}
\def\BbbZ{{\Bbb Z}}
\def\BbbC{{\Bbb C}}

\def\casehalf{{\case{1}{2}}}

\def\morsef{{f}}

\def\so{{\hbox{${\rm SO}(2,1)$}}}

\def\eps{\epsilon}

\def\Re{{\rm Re}}
\def\Im{{\rm Im}}


\preprint{\vbox{\baselineskip=12pt
\rightline{SU--GP--95--5--1}
\rightline{WISC--MILW--95--TH--16}
\rightline{PP96--40}
\rightline{gr-qc/9511023}}}

\title{Complex actions in two-dimensional topology change}
\author{Jorma Louko\footnote{On leave of absence from
Department of Physics,
University of Helsinki.
Electronic address: louko@wam.umd.edu
}}
\address{
Department of Physics,
University of
Wisconsin--Milwaukee,
\\
P.O.\ Box 413,
Milwaukee, Wisconsin 53201, USA
\\
and
\\
Department of Physics,
University of Maryland,
\\
College Park, Maryland 20742-4111, USA\footnote{Present address.}
}
\author{Rafael D. Sorkin\footnote{Electronic address:
rdsorkin@mailbox.syr.edu
}}
\address{
Department of Physics,
Syracuse University,
Syracuse, New York 13244--1130,
USA
\\
and
\\
Instituto de Ciencias Nucleares, UNAM,
A.~Postal 70-543, D.F. 04510, Mexico
}
\date{Revised version, September 1996.
Published in {\it Class.\ Quantum Grav.\ \bf14} (1997) 179--203}
\maketitle

\begin{abstract}%
We investigate topology change in (1+1) dimensions
by analyzing the scalar-curvature action $\casehalf \int R \, dV$ at the
points of metric-degeneration that (with minor exceptions) any
nontrivial Lorentzian cobordism necessarily possesses.  In two dimensions
any cobordism can be built up as a combination of only two elementary
types, the ``yarmulke'' and the ``trousers.''  For each of these
elementary cobordisms, we consider a family of Morse-theory inspired
Lorentzian metrics that vanish smoothly at a single point, resulting in
a conical-type singularity there.  In the yarmulke case, the
distinguished point is analogous to a cosmological initial (or final)
singularity, with the spacetime as a whole being obtained from one causal
region of Misner space by adjoining a single point.  In the trousers
case, the distinguished point is a ``crotch singularity'' that signals a
change in the spacetime topology  (this being also the fundamental vertex of
string theory, if one makes that interpretation). We regularize the metrics
by adding a small imaginary part whose sign is fixed to be positive by the
condition that it lead to a convergent scalar field path integral on the
regularized spacetime.  As the regulator is removed, the scalar density
$\casehalf \sqrt{-g}\,R$ approaches a delta-function whose strength is
complex: for the yarmulke family the strength is $\beta -2\pi i$, where
$\beta$ is the rapidity parameter of the associated Misner space; for the
trousers family it is simply $+2\pi i$.  This implies that in the path
integral over spacetime metrics for Einstein gravity in three or more
spacetime dimensions, topology change via a crotch singularity is
exponentially suppressed, whereas appearance or disappearance of a universe
via a yarmulke singularity is exponentially enhanced.
We also contrast these results with the situation in a
vielbein-cum-connection formulation of Einstein gravity.
\end{abstract}

\narrowtext

\section{Introduction}
\label{sec:intro}

General relativity in $D\ge3$ spacetime dimensions is
most commonly formulated as a generally covariant, nonlinear field theory
for the spacetime metric \cite{MTW,haw-ell,wald}.  One
starts by assuming the (Lorentzian) metric $g_{ab}$ to be smooth (say,
$C^2$ or $C^\infty$) and invertible.  The Riemann tensor $R^a{}_{bcd}$ and
the Einstein tensor $G_{ab}=R_{ab}-\case{1}{2}Rg_{ab}$ are then well
defined, and the gravitational field equations (with matter) read
\begin{equation}
G_{ab} = \kappa T_{ab}
\ ,
\label{EE}
\end{equation}
where $T_{ab}$ is the
stress-energy tensor of the matter and $\kappa$ is
the gravitational constant.\footnote%
{We use units in which $\hbar=c=1$, but we keep the gravitational
constant~$\kappa$. In four dimensions $\kappa=8\pi G$, where $G$ is
Newton's constant.}
Equivalently, the classical
solutions can be defined as the critical points of
an action functional whose gravitational part reads
\begin{equation}
S = {1 \over 2 \kappa} \int d^Dx \sqrt{-g}\,R
\ \ + \ \hbox{(boundary terms)}
\ ,
\label{action}
\end{equation}
where the variation is taken within smooth invertible metrics (and smooth
matter fields) with appropriate boundary conditions.  A variational
formulation is particularly natural if one wishes to view the theory as the
classical limit of a quantum theory of gravity, where it is not the field
equations, but rather the action that plays the central role.

However, the assumption of a smooth, invertible metric is too strong to
accommodate certain situations of physical interest. One such situation,
by now well understood, arises with an idealized matter source whose
$T_{ab}$ is a distribution concentrated on a hypersurface of codimension
one. The metric is then invertible and $C^0$ everywhere, and smooth
outside the timelike $(D-1)$-dimensional world hypersurface of the
matter, but it fails to be $C^1$ across this hypersurface. For
$D=4$, this situation reduces to the familiar case of an infinitely thin
matter shell. The Einstein equations (\ref{EE}) for such spacetimes can
be given a distributional interpretation \cite{GT} that is
equivalent to Israel's junction condition formalism \cite{israel1},
and also the Einstein-Hilbert action (\ref{action}) can be readily
generalized to such spacetimes in a manner that reproduces the junction
condition formalism from a variational principle \cite{haylo}.  These
spacetimes belong to the wider class of invertible but nonsmooth metrics
for which Geroch and Traschen showed that the Einstein equations
(\ref{EE}) have a well-defined distributional interpretation \cite{GT}.
There has also been interest in metrics that incorporate a signature
change on a surface of codimension~1; see for example Refs.\
\cite{sakharov,draymatu1,ellis+,ellis-sigchange,%
hayward1,koss-kriele,draymatu2,hellaby-dray,embacher} and the references
therein.

The main purpose of the present paper is to extend the definition of the
Einstein-Hilbert action~(\ref{action}) to incorporate certain metrics that
are smooth but not everywhere invertible.  The prime motivation for
considering such metrics comes from topology change.  The requirement of a
nondegenerate Lorentzian metric renders topology change in two spacetime
dimensions essentially impossible, while in higher dimensions this
requirement leads to causality violations \cite{raf-vict,borde}, or to
apparently unwarranted restrictions on the allowed transitions, such as the
exclusion of Kaluza-Klein monopole pair production \cite{TMC1,TMC2}.
However, there exists a class of metrics for which the regularity
assumptions are relaxed in a relatively mild manner, but one still broad
enough that metrics of the resulting type can exist on any (smooth)
cobordism in any dimension.  Moreover, these metrics respect causality
(they are compatible with a global causal ordering of the points of the
manifold), and they are free of any other apparent pathology.

The inclusion of these metrics, in which the assumption of global
invertibility has been relaxed, renders topology change
{\em{}kinematically\/} possible, and the issue of topology change becomes a
question about the {\em dynamics\/} of the theory.  Within path-integral
quantization, one must ask how such metrics contribute to the gravitational
path integrals that give transition amplitudes for topology change.  At the
classical level, one can ask whether such metrics are (in an appropriate
sense) critical points of the generalized Einstein-Hilbert action, this
being the criterion for them to make a non-negligible contribution to the
quantum amplitude in the classical limit.

The metrics discussed in \cite{raf-vict,borde} fail to be invertible only
at isolated degenerate {\em points\/}.  In this paper we will consider the
two-dimensional special case of these metrics, as well as
higher-dimensional metrics that locally look like products of the
two-dimensional metrics with a flat metric. Thus, a characteristic
feature of our metrics will be that the degeneracy will be concentrated
on a submanifold of codimension two.  Although the metric itself will
be smooth everywhere, the geometry it describes can be understood as
having a conical-type singularity at the submanifold of degeneracy, and
we shall refer to the locus of degeneracy in this sense as a
singularity.  Singularities of exactly this type appear also in Regge
calculus, as the support of the spacetime curvature
\cite{sorkin-regge}. We note that degeneration of the metric at
submanifolds of positive codimension can be viewed as a particularly
mild form of singularity, if one reflects that the path-integral
measure is expected to be supported entirely on non-smooth
configurations.

When the conically singular submanifold is timelike, or when the whole
spacetime is taken to have Euclidean signature, it is known that the
spacetime can be understood as having a distributional curvature at the
singularity, in a sense that we shall discuss in more detail below (see
also Refs.\
\cite{sorkin-regge,linet,vickers1,vickers2,tod,fursaev,clar-vick-wil}):
the Ricci scalar density has at the singularity a delta-function whose
strength is twice the defect angle. This is the case with
the well-known spacetimes of idealized cosmic strings for $D=4$
\cite{vilenkin,gott,hiscock}, and spacetimes with massive point particles
for $D=3$ \cite{deser1,deser2}.  The Einstein-Hilbert action
(\ref{action}) can be readily extended to metrics of this type
\cite{haylo,sorkin-regge,LW1,cartei1,cartei2,teitel-ext}, with the
result that the singular spacetimes are not critical points of the
action. In the vacuum theory defined in terms of the Einstein-Hilbert
variational principle thus generalized, one therefore concludes that these
singular spacetimes are not classical solutions, but one also finds
that the degrees of freedom associated with the singularity may have
significance when one quantizes the theory by path-integral methods
\cite{haylo,sorkin-regge,LW1,cartei1,cartei2,teitel-ext}.  Note,
however, that extending the variational principle or the field equations
to include an idealized matter source that would make these singular
spacetimes classical solutions has proved difficult
\cite{GT,futamase}.  For example, for an idealized cosmic string in
$D=4$, with $T_{ab}$ a delta-function concentrated on the world sheet of
the string, there is no unambiguous relation between the defect angle and
the linear mass density of the string \cite{futamase}.

Our main aim is to perform a similar analysis in the case where the
spacetime is Lorentzian and the conically singular submanifold is
spacelike.  As in the Riemannian case, we will find that conically singular
metrics are not extrema of the action.  But we also find a surprising new
feature: the action itself is no longer real but complex.

To begin, we consider two one-parameter families of Morse theory
inspired metrics on~$\BbbR^2$. The metrics are everywhere $C^\infty$ (and
even analytic), and they turn out to be flat
with Lorentz signature everywhere
except at a single point, at which they vanish. The metrics of the
first family describe ``yarmulke" spacetimes that can be obtained from
one causal domain of Misner space \cite{haw-ell} by adjoining a single
point, the singularity at the added point being analogous to a
cosmological initial (or final) singularity.  The metrics of the second
family describe a Lorentzian ``trousers" spacetime near the ``crotch
singularity'' \cite{anderson-dewitt,harris-dray,dray+,raf-fluct,horo}.

We wish to give a meaning to the action, and secondarily the
Ricci scalar density, on these two-dimensional spacetimes.  To do so, we
regularize the metrics by adding a small {\em positive\/} imaginary part,
the sign of the regulator following from the requirement that a scalar
field path integral on the regularized spacetime should be (formally)
convergent. Upon taking the regulator to zero, we find that the scalar
density $\casehalf \sqrt{-g}\,R$ converges to a delta-function at the
singularity.  The strength of the delta-function turns out to be complex:
$\beta -2\pi i$ for the yarmulke singularity, where $\beta$ is the
rapidity parameter of the Misner space, and $+2\pi i$ for the crotch
singularity. This agrees with results obtained from a Regge calculus
description of these singular spacetimes
\cite{raf-vict} (see also the appendix of Ref.\ \cite{sorkin-regge}).

Turning to $D\ge3$ dimensions, we then consider spacetimes that are
products of our conical two-dimensional spacetimes
with a $(D-2)$-dimensional compact flat space. If we take the
Einstein-Hilbert action for such
spacetimes to be defined via the same limiting process, it is
immediately seen that the action is not stationary under
variations in the $(D-2)$-volume, and we shall argue that this
non-stationarity remains even after the global product form of the metric
is relaxed and the variations become properly localized. In the theory
that is defined in terms of the Einstein-Hilbert action thus extended, our
singular spacetimes are therefore not classical solutions: topology change
via the yarmulke and trousers cobordisms is classically forbidden.

Within a path-integral quantized theory, the question of topology
change via our metrics amounts to asking how these metrics contribute to
the gravitational path integral. The above results imply that the
contribution from the trousers spacetimes is {\em suppressed\/} by an
exponential factor, over and above what one expects from the fact that
these spacetimes are not stationary points of the action.  Topology
change via the trousers cobordism is therefore exponentially suppressed;
in $3+1$ dimensions the suppression factor is~$\exp(-A/4G)$, where $A$ is
the area of the two flat dimensions. In contrast, the contribution from
the yarmulke spacetimes is {\em enhanced\/} by the same exponential
factor.  One might take this as evidence for an exponential enhancement of
the creation (or annihilation) of a component of spacetime via the
yarmulke cobordism.  However, one would need to balance such an
exponential enhancement against the suppression coming from the fact that
the yarmulke spacetimes are not stationary points of the action.

These conclusions in $D\ge3$ dimensions also provide an interesting
contrast to the  vielbein-cum-connection formulation of Einstein gravity,
within which the trousers spacetime solves the field equations, and even
appears to represent a stationary point of the action in the relevant
sense.  If this is indeed so, then it would seem that the metric and
vielbein-connection formulations must be regarded as physically
inequivalent.

The paper is organized as follows. In Section \ref{sec:regularized} we
introduce the two families of regularized metrics on~$\BbbR^2$, discuss
their properties, and compute the distributional limit of the Ricci scalar
density as the regulator is taken to zero. For the sake of comparison, we
carry the analysis through in both the Lorentzian and Euclidean cases;
in the Euclidean case the regulator can be taken real, and we
reproduce the familiar result for a Euclidean conical singularity. In
Sections \ref{sec:gauss-bonnet}--\ref{sec:imag-sign} we shall present
evidence to the effect that our results are independent of the details of
the regularization, and that they should even be insensitive to
the choice of the differentiable structure at the singularity: one
set of arguments relies on a suitably generalized Gauss-Bonnet theorem,
another on Regge calculus.  Section \ref{sec:discussion} contains a
summary and discussion.
Appendix \ref{app:dcon} compares our method to the vielbein-cum-connection
formulation for the $(2+1)$-dimensional spacetime that is the product of
the time-axis with the ``double cover" conical metric.  Appendix
\ref{app:hourglass} presents a similar comparison for the ``hourglass"
spacetime of Refs.\ \cite{horoPLB,horoMG6}.

\section{Regularized metrics and the Ricci scalar density}
\label{sec:regularized}

As mentioned in the Introduction, a primary reason for our interest in
conical-type singularities is their role in two-dimensional topology
change.  In this context they fit into a general scheme for endowing any
(smooth) topological cobordism\footnote%
{By a topological cobordism, we just mean a compact (say) manifold with
boundary, regarded as mediating a transition between different spacetime
topologies.}
with a Lorentzian metric.  Although a globally regular Lorentzian metric
will exist in some cobordisms, it will not exist in general, and when it
does exist, it necessarily contains time loops (assuming a time-oriented
metric on a compact, non-product cobordism with spacelike boundaries).
Even accepting these time loops, one still cannot find Lorentzian metrics
for certain cases where, on physical grounds, one would want them to exist,
such as on the natural cobordism for mediating the pair creation of
Kaluza-Klein monopoles in five dimensions.
For a general discussion, see for example Refs.\
\cite{TMC1,TMC2,GHselection} and the references therein.
In contrast, there always exists a Morse function \cite{bott-morse}
$\morsef$ on any cobordism, as well as a positive-definite, Riemannian
metric~$h_{ab}$.  {}From such a pair one can construct a metric that is
almost Lorentzian (in the sense of being smooth with Lorentzian signature
everywhere except at the critical points of~$\morsef$), and further has  a
well-defined causal structure for which $\morsef$ furnishes a ``time
function'' \cite{raf-vict,borde}.  The resulting geometry can thus be taken
to contribute to the path integral amplitude for the topology change which
the cobordism mediates. Specifically, the metric in question has the form
\begin{equation}
g_{ab} =
\left( \partial_c \morsef \right)
\left( \partial_d \morsef \right)
h^{cd}
h_{ab}
- \zeta
\left( \partial_a \morsef \right)
\left( \partial_b \morsef \right),
\label{morse}
\end{equation}
where the parameter $\zeta$ must satisfy $\zeta>1$ in order that the
signature be Lorentzian.  The metrics we will study in the present paper
belong to this family with $h_{ab}$ chosen as flat and with $\morsef$ a
quadratic polynomial in Cartesian coordinates for~$h_{ab}$.  By a
remarkable accident, (\ref{morse}) then turns out (in two dimensions) to
be flat everywhere except at the origin \cite{luca}.

Let $(x,y)$ be a global coordinate chart on~$\BbbR^2$. We shall
consider on $\BbbR^2$ the metric \cite{raf-vict,borde}
\begin{equation}
g_{ab} =
\left( \partial_c \morsef \right)
\left( \partial_d \morsef \right)
\delta^{cd}
\delta_{ab}
- \zeta
\left( \partial_a \morsef \right)
\left( \partial_b \morsef \right)
+ \gamma \delta_{ab}
\ ,
\label{metric1}
\end{equation}
where $\morsef=\casehalf \left( x^2 + \epsilon y^2 \right)$ is the Morse
function, $\epsilon=\pm1$ is essentially the Morse index of~$\morsef$, and
$\zeta$ and $\gamma$ are parameters.  We take $\gamma\in\BbbC$, and
$\zeta\in\BbbR\setminus \{1\}$. In the chart $(x,y)$ we have then
\begin{equation}
ds^2 = g_{ab} dx^a dx^b
= \left( x^2 + y^2 + \gamma \right)
\left( dx^2 + dy^2 \right)
- \zeta {\left( x dx + \epsilon y dy \right)}^2
\ .
\label{metric2}
\end{equation}
This metric is $C^\infty$ (and even analytic) in the differentiable
structure determined by the chart~$(x,y)$, but it need not be everywhere
invertible. Our aim is to investigate the limit $\gamma=0$, in which the
metric is invertible everywhere except at the point $x=y=0$. We shall treat
$\gamma$ as a regulator, chosen to make the metric invertible, and we will
examine the curvature in the limit $\gamma\to0$.
There are four qualitatively different cases depending on $\epsilon$ and
the sign of $\zeta-1$.  We shall devote a separate subsection to each case.
\subsection{General Euclidean conical singularity}
\label{subsec:+xismall}

We take first $\epsilon=1$, $\zeta<1$.
When $\gamma=0$, the metric (\ref{metric2}) has signature $(++)$ everywhere
except at $x=y=0$.  To make the geometry more transparent, we perform for
$x^2+y^2>0$ the coordinate transformation
\begin{equation}
\begin{array}{rcl}
  x &=& \sqrt{r} \cos\varphi
  \ ,
  \\
  y &=& \sqrt{r} \sin\varphi
  \ ,
\end{array}
\label{transf1}
\end{equation}
where $r>0$. We understand $\varphi$ to be periodically identified with
period~$2\pi$, $(r,\varphi)\sim(r,\varphi+2\pi)$. The metric reads then
\begin{equation}
  ds^2 = \case{1}{4} (1-\zeta) dr^2 + r^2 d\varphi^2
  \ ,
  \label{baremetric}
\end{equation}
which is recognizable as the metric of a cone with defect angle
$2\pi\left( 1 - 2 {(1-\zeta)}^{-1/2} \right)$.
Note that when $\zeta=-3$ the defect angle vanishes
and the cone reduces to a plane.

We now take the regulator $\gamma$ to be positive. The metric
(\ref{metric2}) has then signature $(++)$ everywhere, including the origin.
With the notation
\begin{mathletters}
\begin{eqnarray}
  N &=& \sqrt{(1-\zeta)\rho^2 + \gamma}
  \ ,
  \\
  a &=& \rho \sqrt{\rho^2 + \gamma}
  \ ,
\end{eqnarray}
\end{mathletters}%
where $\rho = \sqrt{x^2+y^2} = \sqrt{r}$, the Ricci scalar takes on the
appearance
\begin{equation}
   R = - {2\over aN} {\left( {a' \over N} \right)}'
   \ ,
\end{equation}
where the prime denotes derivative with respect to~$\rho$.  The volume
element of the metric is given by
\begin{equation}
  \sqrt{g}\,d^2x
  =
  \rho^{-1}aN \, dxdy
  \ .
\end{equation}

We wish to examine the scalar density $\casehalf \sqrt{g}\,R$ in the limit
$\gamma\to0$.  For this purpose, let $\Phi(x,y)$ be a test function on
$\BbbR^2$ \cite{reed-simon},\footnote{The distinction between test functions
of rapid decrease and test functions of compact support will be irrelevant
here.}  and consider the integral of the density $\casehalf \sqrt{g}\,R
\Phi$.  Using the polar coordinates
\begin{equation}
  \begin{array}{rcl}
  x &=& \rho \cos\varphi
  \ ,
  \\
  y &=& \rho \sin\varphi
  \ ,
  \end{array}
  \label{transf2}
\end{equation}
and writing
\begin{equation}
  {\overline\Phi}(\rho) =
  {(2\pi)}^{-1} \int_0^{2\pi} d\varphi \, \Phi(\rho
  \cos\varphi,\rho\sin\varphi)
  \ ,
  \label{Phibar}
\end{equation}
we obtain
\begin{eqnarray}
  \casehalf \int_{\BbbR^2} d^2x \, \sqrt{g}\, R\Phi
  &=&
  - \int_{\BbbR^2} dx dy \, {1\over \rho}
  {\left( {a' \over N} \right)}'
  \Phi
  \nonumber
  \\
  \noalign{\smallskip}
  &=&
  -2\pi \int_0^\infty d\rho {\left( {a' \over N} \right)}'
  {\overline \Phi}(\rho)
  \nonumber
  \\
  \noalign{\smallskip}
  &=&
  -2\pi \int_0^\infty d\rho {\left( {a' \over N} - {2\over
  \sqrt{1-\zeta}} \right)}' {\overline \Phi}(\rho)
  \nonumber
  \\
  \noalign{\smallskip}
  &=&
  2\pi \left( 1 - {2\over \sqrt{1-\zeta}} \right) {\overline \Phi}(0)
  \;
  + 2\pi \int_0^\infty d\rho {\left( {a' \over N} - {2\over
  \sqrt{1-\zeta}} \right)} {\overline \Phi}'(\rho)
  \ ,
  \label{osint-eucl}
\end{eqnarray}
where in the last step we have integrated by parts and used the fact
that $a'/N \to 1$ as $\rho\to0$.
Note that ${\overline \Phi}'(\rho)$ vanishes at $\rho=0$.
Now, in the limit $\gamma\to0$, the
integral
term
on the last line of (\ref{osint-eucl}) vanishes by
dominated convergence. As ${\overline \Phi}(0)=\Phi(0,0)$, this means that
the $\gamma\to0$ limit\footnote%
{Strictly speaking, what we have established is that the $\gamma\to0$ limit
exists with respect to the topology of pointwise convergence (``weak
topology'').  For many purposes a finer topology on the space of
distributions is desirable, such as that treated in Chapter III of
Ref.\ \cite{Schwartz}.  Convergence with respect to the latter topology
actually follows from Theorem XIII of that chapter, but as a rule we will
ignore such distinctions herein, as our primary concern is limited to
the evaluation of the action integral~(\ref{action}). }
of the density $\casehalf \sqrt{g}\, R$ is the distribution
\begin{equation}
\casehalf \sqrt{g}\, R =
2\pi \left( 1 - {2\over \sqrt{1-\zeta}} \right) \delta_2(x,y)
\ ,
\label{Rdens-e}
\end{equation}
where $\delta_2(x,y)$ stands for the two-dimensional delta-function
concentrated at $x=y=0$. The coefficient of the delta-function in
(\ref{Rdens-e}) is precisely the defect angle. This is the well-known
result for a conical singularity \cite{sorkin-regge}.

Note that in (\ref{Rdens-e}) the left hand side is understood as a single
entity: we are not attempting here to give a distributional meaning to the
Ricci scalar $R$ as such. We shall return to this issue in
Sections \ref{sec:gauss-bonnet} and~\ref{sec:curvature}.

\subsection{Lorentzian yarmulke singularity}
\label{subsec:+xilarge}

We take next $\epsilon=1$, $\zeta>1$.

When $\gamma=0$, the metric (\ref{metric2}) now has signature $(-+)$
everywhere except at $x=y=0$. The coordinate transformation (\ref{transf1})
brings the metric for $x^2+y^2>0$ again to the form~(\ref{baremetric}). If
$\varphi$ were not periodic, the transformation
\begin{equation}
 \begin{array}{rcl}
  T &=& \casehalf { \displaystyle {{(\zeta-1)}^{1/2} r \cosh\left( 2
  {(\zeta-1)}^{-1/2} \varphi  \right) }}
  \ ,
  \\
  \noalign{\medskip}
  X &=& \casehalf { \displaystyle {{(\zeta-1)}^{1/2} r \sinh\left( 2
  {(\zeta-1)}^{-1/2} \varphi  \right) }}
 \end{array}
\end{equation}
would bring this metric into the explicit Minkowski form
\begin{equation}
  ds^2 = - dT^2 + dX^2
  \ ,
  \label{minkmetric}
\end{equation}
where the range of the coordinates would be $T>|X|$. The periodicity of
$\varphi$ therefore means that one has to take the quotient of the domain
$T>|X|$ in (\ref{minkmetric}) with respect to a boost whose rapidity
parameter is $4\pi {(\zeta-1)}^{-1/2}$.  The resulting spacetime is known
as (one of) the causal region(s) of Misner space \cite{haw-ell}.  The full
spacetime described by the metric (\ref{metric2}) with $\gamma=0$ is thus
obtained by adding a single point to one
causal region
of Misner space. If the coordinate $T$ is taken to increase towards the
future (respectively past), the added point is to the past (future) of
every other point, and it can be regarded as analogous to a cosmological
initial (final) singularity.  Continuing the sartorial imagery of
Refs.\
\cite{anderson-dewitt,harris-dray,dray+}, we shall refer to this spacetime
as a yarmulke spacetime, and to the critical point at $x=y=0$ as a yarmulke
singularity.

We now take the regulator $\gamma$ to be purely imaginary:
$\gamma=\pm i\sigma$, where $\sigma>0$. The metric (\ref{metric2}) is
then complex, but everywhere non-degenerate. Writing
\begin{mathletters}
 \label{lor-Ma}
  \begin{eqnarray}
    M &=& \sqrt{\rho^2 (\zeta-1) - \gamma}
    \ ,
    \\
    a &=& \rho \sqrt{\rho^2 + \gamma}
    \ ,
  \end{eqnarray}
\end{mathletters}%
where $\rho = \sqrt{x^2+y^2}$ as before, lends the Ricci scalar the form
\begin{equation}
  R = + {2\over aM} {\left( {a' \over M} \right)}'
  \ ,
  \label{ricci-lor}
\end{equation}
and the volume element is given by
\begin{equation}
  \sqrt{-g}\,d^2x
  =
  \rho^{-1}aM \, dxdy
  \ .
  \label{dens-lor}
\end{equation}
Notice that this volume element would have vanished for some positive
$\rho$ if we had tried to take $\gamma$ real.
We wish to examine the scalar density $\casehalf\sqrt{-g}\,R$.

Although the expression (\ref{ricci-lor}) for $R$ does not depend on the
branches chosen for the square roots in~(\ref{lor-Ma}), the expression
(\ref{dens-lor}) for $\sqrt{-g}$ does. In order to guarantee that the
$\gamma\to0$ limit of $\sqrt{-g}$ is positive for $\rho>0$, we choose the
real parts of the square roots in (\ref{lor-Ma}) to be positive at large
values of~$\rho$. Integrating $\casehalf \sqrt{-g}\,R$ against a test
function $\Phi(x,y)$ and proceeding as in the previous subsection yields
then
\begin{equation}
  \casehalf \int_{\BbbR^2} d^2x \, \sqrt{-g}\, R\Phi
  =
  2\pi \left( {2\over \sqrt{\zeta-1}} \mp i \right) {\overline \Phi}(0)
  \; - 2\pi \int_0^\infty d\rho {\left( {a' \over M} - {2\over
  \sqrt{\zeta-1}} \right)} {\overline \Phi}'(\rho)
  \ .
  \label{osint-lor}
\end{equation}
In the limit $\gamma\to0$, the integral
term
on the right hand side of (\ref{osint-lor}) vanishes by dominated
convergence. Therefore, the
$\gamma\to0$ limit of $\casehalf \sqrt{-g}\,R$ is again a two-dimensional
delta-function at $x=y=0$:
\begin{equation}
  \casehalf \sqrt{-g}\, R =
  2\pi \left( {2\over \sqrt{\zeta-1}} \mp i \right)
  \delta_2(x,y)
  \ ,
\label{Rdens-lor}
\end{equation}
where the sign $\mp$ corresponds to the sign in $\gamma=\pm{i}\sigma$.

The strength of the delta-function is now complex. Its real part is equal
to the rapidity parameter of the Misner space, but there is also an
imaginary
     piece~$\mp 2\pi i$, which
is entirely independent of the rapidity
parameter.  The sign of this imaginary piece depends on the sign of our
imaginary regulator $\gamma=\pm{i}\sigma$.  We can fix this sign by
requiring that the action
\begin{equation}
  S_\phi =
     - \case{1}{2} \int \sqrt{-g}\,d^2x \,
      g^{ab} (\partial_a\phi) (\partial_b\phi)         \label{Sphi}
\end{equation}
for a massless scalar field have a positive imaginary part for real-valued
$\phi$ on the regularized spacetime, this being the (formal) condition of
convergence for path integrals of the form $\int D\phi \exp\left( iS_\phi
\right)$ on the regularized spacetime.  It is straightforward to verify that
satisfaction of this condition is equivalent to the choice
$\gamma=+i\sigma$, which yields the upper sign in~(\ref{Rdens-lor}):
\begin{equation}
  \casehalf \sqrt{-g}\, R =
   2\pi \left( {2\over \sqrt{\zeta-1}} - i \right) \delta_2(x,y) \ .
  \label{Rdens-lor+}
\end{equation}
Notice that with the choice $\Im\gamma > 0$, the regularized metric
satisfies the general condition,
\begin{equation}
                {\rm Im\,}(g_{ab})>0 \ ,  \label{sharp}
\end{equation}
This puts our choice of sign in a broader context, as will be discussed in
Section~\ref{sec:imag-sign}.

\subsection{Euclidean double cover conical singularity}
\label{subsec:-xismall}

We take now $\epsilon=-1$, $\zeta<1$.

When $\gamma=0$, the metric (\ref{metric2}) has signature $(++)$ everywhere
except at $x=y=0$.  To understand the geometry, we perform the local
coordinate transformation
\begin{equation}
 \begin{array}{rcl}
  u &=&  x^2 - y^2
  \ ,
  \\
  v &=& 2xy
  \ ,
 \end{array}
 \label{esquare}
\end{equation}
which brings the metric to the explicitly flat form
\begin{equation}
  ds^2 = \case{1}{4} (1-\zeta) du^2 + \case{1}{4} dv^2
  \ .
  \label{-baremetric}
\end{equation}
Globally, the transformation (\ref{esquare}) is the squaring map
in the complex plane, $u+iv={(x+iy)}^2$. This means that for $x^2+y^2>0$,
the metric (\ref{metric2}) describes the double cover of the punctured flat
Euclidean plane. In other words, the metric (\ref{metric2}) describes a
cone with defect angle $-2\pi$.  Note that the defect angle is negative and
independent of~$\zeta$, so that all the metrics (\ref{-baremetric})
describe in some sense the same geometry. Ramifications of this
fact will be discussed in Sections
\ref{sec:curvature} and~\ref{sec:discussion}.

We now take the regulator $\gamma$ to be positive, making the metric
everywhere positive definite.  Writing
\begin{mathletters}
 \label{AB}
 \begin{eqnarray}
  A &=& \rho^2 + \gamma
  \ ,
  \label{Adef}
  \\
  B &=& (1-\zeta)\rho^2 + \gamma
  \ ,
 \end{eqnarray}
\end{mathletters}%
where $\rho = \sqrt{x^2+y^2}$ as before, brings the Ricci scalar to the
form
\begin{equation}
  R = - 2\gamma A^{-2} B^{-2}
  \left( 2 (1-\zeta)\rho^2 +
  (2-\zeta)\gamma
  \right)
  \ ,
  \label{-ricci}
\end{equation}
and the volume element to the form
\begin{equation}
  \sqrt{g}\,d^2x
  =
  \sqrt{AB} \, dxdy
  \ .
  \label{dens-e2}
\end{equation}

We wish to consider the scalar density $\casehalf \sqrt{g}\, R$.  Let
$\Phi(x,y)$ be a test function.  Employing the polar
coordinates~(\ref{transf2}), writing $\rho^2=r$, and making the
definition~(\ref{Phibar}), we obtain
\begin{eqnarray}
\casehalf \int_{\BbbR^2} d^2x \, \sqrt{g}\, R\Phi
&=&
-\gamma \int_{\BbbR^2} dx dy \,
{(AB)}^{-3/2}
\left( 2 (1-\zeta)\rho^2 + (2-\zeta)\gamma \right)
\Phi
\nonumber
\\
\noalign{\smallskip}
&=&
-2\pi \gamma \int_0^\infty d\rho \, \rho
{(AB)}^{-3/2}
\left( 2 (1-\zeta)\rho^2 + (2-\zeta)\gamma \right)
{\overline \Phi}
\nonumber
\\
\noalign{\smallskip}
&=&
-\pi \gamma \int_0^\infty dr \,
{(AB)}^{-3/2}
\left( 2 (1-\zeta)r + (2-\zeta)\gamma \right)
{\overline \Phi}
\nonumber
\\
\noalign{\smallskip}
&=&
2\pi \gamma \int_0^\infty dr \,
{\overline \Phi} \,
{d \over dr} \! \left[
{(AB)}^{-1/2} \right]
\nonumber
\\
\noalign{\smallskip}
&=&
-2\pi {\overline \Phi} (0)
\; - 2\pi \int_0^\infty dr \,
\gamma {(AB)}^{-1/2}
\, {d {\overline \Phi}
\over dr}
\ ,
\label{-osint-eucl}
\end{eqnarray}
where in the last step we have integrated by parts and used the fact that
${(AB)}^{1/2} \to \gamma$ as $r\to0$.
Note that $d{\overline \Phi}/dr$ is finite at $r=0$.
In the limit $\gamma\to0$, the
integral
term
on the last line of (\ref{-osint-eucl}) vanishes by dominated
convergence.  Hence, the $\gamma\to0$ limit of $\casehalf\sqrt{g}\, R$ is a
two-dimensional delta-function at $x=y=0$,
\begin{equation}
\casehalf \sqrt{g}\, R =
-2\pi \delta_2(x,y)
\ ,
\label{Rdens-dcover}
\end{equation}
the strength of the delta-function again being precisely the defect angle.
The result therefore agrees with that obtained in
subsection~\ref{subsec:+xismall}, whose special case $\zeta=0$ corresponds
to the entire family of  metrics of the present
subsection.

\subsection{Lorentzian crotch singularity}
\label{subsec:-xilarge}

Finally we take $\epsilon=-1$, $\zeta>1$.

When $\gamma=0$, the metric (\ref{metric2}) has signature $(-+)$ everywhere
except at $x=y=0$.  The coordinate transformation (\ref{esquare}) again
brings the metric to the form~(\ref{-baremetric}), which is now explicitly
flat Lorentzian.  This means that for $x^2+y^2>0$, the metric
(\ref{metric2}) describes the double cover of punctured two-dimensional
Minkowski space.  The geometry near the origin is therefore that of the
flat $(1+1)$-dimensional trousers spacetime
\cite{raf-vict,anderson-dewitt,harris-dray,dray+,raf-fluct,horo}.
To construct a spacetime that was globally a trousers, one would take
the star-shaped domain defined by the inequalities
\begin{equation}
 \begin{array}{lll}
                    |x| \, y   \le a &&\hbox{for $y\ge0$}\ ,
  \\
  \noalign{\smallskip}
                    |x| \, y  \ge - b &&\hbox{for $y\le0$}\ ,
 \end{array}
\end{equation}
where $a$ and $b$ are positive constants, and on the boundary of this
domain one would perform the identifications
\begin{equation}
\begin{array}{lll}
(a/y,y) \sim (-a/y,y) &&\hbox{for $y>0$}\ ,
\\
\noalign{\smallskip}
(b/y,y) \sim (-b/y,y) &&\hbox{for $y<0$}\ .
\end{array}
\label{trousers-identifications}
\end{equation}
The identified lines become the outer seams of the trousers.  The two
legs are at $y\to\infty$ and $y\to-\infty$, with geodesic circumferences
$a$ and $b$ respectively, and the waist is at $|x|\to\infty$, with
geodesic circumference $a+b$.

As in subsection~\ref{subsec:+xilarge}, a real regulator cannot yield an
invertible metric.  We must therefore again take $\gamma$ to be complex,
and again we
choose it purely imaginary: $\gamma=\pm i\sigma$, where
$\sigma>0$.  The metric is then complex, but everywhere non-degenerate.
The Ricci scalar is given by~(\ref{-ricci}), with $A$ and $B$ as in
(\ref{AB}), and for the volume element we have now
\begin{equation}
  \sqrt{-g}\,d^2x
  =
  \sqrt{-AB} \, dxdy
  \ .
  \label{dens-lor2}
\end{equation}
To ensure that the $\gamma\to0$ limit of $\sqrt{-g}$ is positive for
$\rho>0$, we choose the branch of the square root in (\ref{dens-lor2}) so
that real part of $\sqrt{-g}$ is positive at large values of~$\rho$.
Integrating $\casehalf \sqrt{-g}\,R$ against a test function $\Phi(x,y)$
and proceeding as in the previous subsection yields
\begin{equation}
  \casehalf \int_{\BbbR^2} d^2x \, \sqrt{-g}\, R\Phi
  =
  \pm 2\pi i {\overline \Phi}(0)
  \; + 2\pi \int_0^\infty dr \,
  \gamma {(-AB)}^{-1/2}
  \, {d {\overline \Phi}
  \over dr}
  \ .
  \label{-osint-lor}
\end{equation}
In the limit $\gamma\to0$, the integral
term
on the right hand side of
(\ref{-osint-lor}) vanishes by dominated convergence.  Therefore, the
$\gamma\to0$ limit of the scalar density $\casehalf \sqrt{-g}\, R$ is
once again a two-dimensional delta-function at $x=y=0$:
\begin{equation}
  \casehalf \sqrt{-g}\, R =
  \pm 2\pi i \, \delta_2(x,y)
  \ .
  \label{Rdens-lor2}
\end{equation}
The coefficient of the delta-function is now purely imaginary, and its sign
depends on the imaginary part of the regularized metric.  Requiring
$\Im(g_{ab})>0$ for the same reasons as before again fixes $\gamma$ to be
positive imaginary, which in turn yields the upper sign
in~(\ref{Rdens-lor2}):
\begin{equation}
  \casehalf \sqrt{-g}\, R =
  2\pi i \, \delta_2(x,y)
  \ .
  \label{Rdens-lor2+}
\end{equation}
Notice that the strength of the
delta-function is independent of the parameter~$\zeta$. As in
subsection~\ref{subsec:-xismall}, this reflects the fact that all
values of $\zeta$ describe, in an appropriate sense, the same geometry.
We shall return to this issue in Sections
\ref{sec:curvature} and~\ref{sec:discussion}.

\section{Action and curvature from the Gauss-Bonnet theorem (or Regge
         calculus)}
\label{sec:gauss-bonnet}

We have seen that when our singular metrics are approached within our
family of regularized metrics, the Ricci scalar density converges to a
delta-function. The strength of this delta-function depends only on the
defect angle in the Euclidean case, and only on the rapidity parameter in
the Lorentzian case. In this section we recall how the Euclidean
results follow from the Gauss-Bonnet theorem, and argue that the
Lorentzian results can also be recovered from a suitable generalization of
the Gauss-Bonnet theorem.

The simplest form of the Gauss-Bonnet theorem states that, given a smooth
positive definite metric on a closed two-dimensional manifold, the
integral of the scalar density
$\casehalf\sqrt{g}\,R$ is a topological invariant independent of the
metric \cite{spivak,allen-weil}:
\begin{equation}
\casehalf \int d^2x \, \sqrt{g}\,R
\; = 2\pi\chi
\ ,
\label{gauss-bonnet}
\end{equation}
where $\chi$ is the Euler number of the manifold. If one requires that
this theorem hold also for positive definite metrics with conical
singularities, it is a matter of simple geometry to see that the
contribution to $\casehalf\sqrt{g}\,R$ at the singularity must contain a
delta-function whose strength is equal to the defect angle. (For example,
for defect angles between $0$ and $2\pi$ one can consider $S^2$ with
the metric consisting of a cone joined in a $C^1$ fashion to a spherical
cap that is larger than a hemisphere.)  If one further requires that
$\casehalf\sqrt{g}\,R$ contain at the singularity no worse
distributions than the delta-function, the result is then entirely
fixed for the curvature, as well as the action.
The requirement that there be no derivatives of delta-functions
at the singularity can be motivated by the fact
that, for our unregularized metrics, the most singular
individual terms in the scalar curvature density diverge like
${(x^2+y^2)}^{-1}$ near the singularity.

The fact that the density $\casehalf \sqrt{g}\,R$ is a total divergence in
two dimensions remains true for complex-valued metrics. The integral of
this density must therefore remain invariant under continuous local
deformations of the metric even when the metric is complex. One therefore
expects that the Euclidean Gauss-Bonnet theorem can be in some suitable
sense analytically continued to complex metrics, and eventually to
Lorentzian or almost Lorentzian metrics.  For a compact manifold, this
would mean continuing (\ref{gauss-bonnet}) to
\begin{equation}
\casehalf \int d^2x \, \sqrt{-g}\,R
\; = -2\pi i\chi
\ .
\label{gauss-bonnet-lor}
\end{equation}
The sign on the right hand side of
(\ref{gauss-bonnet-lor}) follows from adopting the usual direction for
Wick rotation, or equivalently, from our continuation rule~(\ref{sharp}).
Note that (\ref{gauss-bonnet-lor}) is in agreement with the fact that a
closed two-manifold admits a (strictly) Lorentzian metric only for
$\chi=0$; see for example Refs.\
\cite{TMC1,TMC2,reinhart} and the references therein.
Note also that a Gauss-Bonnet theorem in two dimensions is known for
Lorentzian metrics on compact manifolds with a boundary, when this boundary
consists solely of spacelike and timelike segments \cite{jee,birman,law}:
in this theorem, the left hand side of (\ref{gauss-bonnet-lor}) contains
also line integral terms from the boundary segments and corner terms from
the points where the segments meet.

If (\ref{gauss-bonnet-lor}) is assumed to hold for metrics that are
almost Lorentzian,
our results for the scalar density $\casehalf \sqrt{-g}\,R$ at the yarmulke
and crotch singularities become immediate, just as in the Euclidean case.
To see this for the yarmulke singularity,  consider (for example) the
manifold $\BbbR\times S^1$ with the metric
\begin{equation}
ds^2 = - dt^2 + \sin^2 (t) d\theta^2
\ ,
\label{1+1ads}
\end{equation}
where $0<t<\pi$, and $\theta$ is periodic with period~$\beta>0$.  Locally
(\ref{1+1ads}) is just the $(1+1)$-dimensional anti-de~Sitter metric with
$R=-2$. Adding a point at $t=0$ and another point at $t=\pi$ yields the
manifold $S^2$ with two yarmulke singularities, each having rapidity
parameter~$\beta$. The right hand side of (\ref{gauss-bonnet-lor}) is equal
to~$-4\pi i$, and the contribution to the left hand side from the smooth
part is~$-2\beta$: hence $\casehalf \sqrt{-g}\,R$ must have at each
singularity a delta-function with the strength~$\beta-2\pi i$, which is
the result~(\ref{Rdens-lor+}).

To obtain the analogous
result for the crotch singularity, begin with the trousers
spacetime, with the identifications~(\ref{trousers-identifications}), and
close off the waist and each of the legs with a yarmulke consisting of
the half
$0\le{t}\le\casehalf\pi$ of (\ref{1+1ads}) as just described.  The
result is the manifold~$S^2$, with one crotch singularity and three
yarmulke singularities. The nonvanishing volume integrals are essentially
the same as before, and with the result for the yarmulke singularity known,
(\ref{gauss-bonnet-lor}) immediately yields for $\casehalf \sqrt{-g}\,R$ at
the crotch singularity the result~(\ref{Rdens-lor2+}).  Alternatively,
instead of closing off the legs and waist, one can employ the Gauss-Bonnet
theorem for manifolds with boundary \cite{chern}; the result is the same.

The subtlety with the anticipated complex Gauss-Bonnet theorem is that
when the metric is complex, the invariance of $\casehalf\int
d^2x \sqrt{g}\,R$ under continuous deformations of the metric is {\em
not\/} by itself sufficient to fix the value of the integral of this
density in terms of the topology of the manifold: there
exist complex metrics for which neither (\ref{gauss-bonnet}) nor
(\ref{gauss-bonnet-lor}) holds, even after taking into account the
possibility of the global sign ambiguity in
$\sqrt{g}$.\footnote%
{For example, $S^2$ admits smooth invertible complex
metrics for which $\casehalf \int_{S^2} d^2x \sqrt{g}\,R=0$, whereas for
positive definite metrics the Gauss-Bonnet theorem (\ref{gauss-bonnet})
implies $\casehalf \int_{S^2} d^2x \sqrt{g}\,R=4\pi$.  An explicit example
follows.  Define the function $f\colon(0,\pi)\to\BbbC$ by $f(\tau) =
\sin(\tau)\left[1+i\sin^2(\tau)\cos(\tau)\right]$.  Consider the metric
$ds^2 = {\left[f'(\tau)\right]}^2 d\tau^2 + f^2(\tau) d\varphi^2$, where
$f'=df/d\tau$ and $\varphi$ is periodic with period~$2\pi$.  Near $\tau=0$
and $\tau=\pi$ one can introduce new local coordinates (for example,
$x=\tau\cos\varphi$ and $y=\tau\sin\varphi$ near $\tau=0$), in which one
sees that $\tau=0$ and $\tau=\pi$ are just coordinate singularities and
that the metric can be naturally completed into a nondegenerate metric
on~$S^2$.  As the metric is locally related by a complex
diffeomorphism \cite{hall-hartle} to the flat Euclidean metric in polar
coordinates, $ds^2 = dr^2 + r^2 d\varphi^2$, the Riemann tensor vanishes,
and hence $\casehalf \int_{S^2} d^2x \sqrt{g}\,R=0$.  (The total
volume of this metric vanishes,
$\int_{S^2} d^2x \sqrt{g}=0$, but one can easily deform the metric
locally in a
$C^\infty$ fashion to make the volume nonzero while retaining
$\casehalf \int_{S^2} d^2x \sqrt{g}\,R=0$.)}
Nevertheless, it appears plausible that
one can formulate a complex Gauss-Bonnet theorem by placing suitable
restrictions of a topological nature on the complex metric involved.
These restrictions  could then be used to specify a class of
regularizations within which the resulting value for the Ricci scalar
density would agree with ours---something like regularizations for which
the complex metric has positive imaginary part and contains some positive
definite metric in its connected component.  One might also expect to be
able to characterize the appropriate connected component by a ``winding
number'' of the cross-section it represents of the bundle of complex-valued
metrics on the manifold.  We have not attempted to explore this
question in a systematic fashion, however.

Finally, we note that the complex angles that implicitly occur in
Lorentzian Regge calculus provide yet another route to our basic results
for the integrated scalar curvature.  If one subdivides a neighborhood of
the crotch singularity (say) into flat simplicial blocks, then the defect
angle can be computed easily using the ``complex trigonometry'' of the
appendix of Ref.\ \cite{sorkin-regge}. The answer again agrees with
what we have found above, provided one resolves the
complex-conjugation ambiguity analogously.

\section{Tensor densities and distributional curvature}
\label{sec:curvature}

Until now we have concentrated on the Ricci scalar density. In this section
we shall briefly comment on the possibilities for giving a distributional
interpretation to other curvature quantities of interest.

At a general level, we recall that the ordinary delta-function on a
manifold can, by definition, be integrated against a test scalar without
invoking a volume element: in a local coordinate system $(x^1, x^2,
\ldots, x^n)$ whose origin is at the point of support of the
delta-function, the single component of the ordinary delta-function is
just $\delta(x^1)\delta(x^2) \cdots \delta(x^n)$. This means that the
ordinary delta-function should be viewed as a singular {\em scalar
density\/} of weight one, not as a singular function of weight zero.
Indeed, as the ordinary delta-function is a {\em measure\/}, it requires
only $C^0$ test functions, and hence it is  insensitive to the choice of
the differentiable structure (depending only on the manifold's topology).
In contrast, the scalar (or ``covariant") delta-function
${(\pm g)}^{-1/2}\delta$, defined on a (pseudo-)Riemannian manifold, can
be thought of as a singular function of weight zero:\footnote%
{On a manifold that has not been endowed with a volume element it
is not possible to define a unique delta-function of density weight
zero.  The best one can do is define a one-parameter family of
such delta-functions, all of which are scalar multiples of each other,
but a unique choice of normalization requires a volume element.}
it should be integrated against test densities of weight one, and the
definition of such test densities requires knowledge of the differentiable
structure.

Now, we have seen that the Ricci scalar density for our singular metrics
is proportional to the ordinary delta-function on the two-manifold.
Although we worked within a particular set of coordinates, both sides of
this equation are singular densities of weight one, and the equation
therefore must hold in arbitrary coordinates, and with any choice
of the differentiable structure. In effect, we already relied on these
properties in Section \ref{sec:gauss-bonnet} when using the Gauss-Bonnet
theorem to show that the result is independent of the details of the
regularization. What we wish to emphasize here is that the result is
not tied to the differentiable structure defined by our coordinate chart
$(x,y)$.

The reason for making the above point is that the differentiable
structure defined by the chart $(x,y)$ need not always be a natural one
from the viewpoint of the $\gamma=0$ geometry.  For example, when
$\epsilon=1$ and $\zeta=-3$, the defect angle vanishes and the metric
(\ref{metric2}) can be thought of as describing a plane, but the usual
Cartesian coordinates on the plane are then not differentiable functions in
the chart $(x,y)$ at the origin. For $\epsilon=-1$, all values
of $\zeta<1$ give a metric with the defect angle~$-2\pi$, but if one
seriously interprets the metric (\ref{metric2}) with $\gamma=0$ and two
different values of $\zeta<1$ as describing the same geometry in two
different coordinate systems, one finds that the transformation between the
two coordinate systems is continuous but not differentiable at the origin.
Completely analogously, for $\epsilon=-1$, all values
of $\zeta>1$ give a metric with ``the same'' crotch singularity at the
origin, but if one interprets the metric
(\ref{metric2}) with $\gamma=0$ and two different values of $\zeta>1$ as
describing the same geometry in two different coordinate systems, the
coordinate transformation is not differentiable at the origin. None of
this raises concerns about our result for the Ricci scalar density,
however. Indeed, for $\epsilon=-1$, the result obtained in
Section \ref{sec:regularized} was explicitly seen to depended on
$\zeta$ only through whether $\zeta<1$ or $\zeta>1$.

If one attempts to derive from our metrics distributional curvature tensors
(or scalar densities of weight other than one),
the role of the differentiable structure becomes more important.  One might
still expect the Riemann and Ricci curvatures to be concentrated at the
singular point as some sort of  delta-function, but any such relation would
now have to rely on the specification of a reference differentiable
structure, or at least a reference  volume element.  This relative character
of the curvature being defined would make it  more difficult to ascertain
to what extent its distributional limit depended on the particular
regularization chosen.

As an example, consider the Riemann tensor for our metrics. When the metric
is invertible, the independent components of the Riemann tensor
$R^a{}_{bcd}$ in the chart $(x,y)$ take the form
\begin{equation}
  \begin{array}{rcl}
  R^x{}_{xxy} &=&
  \casehalf g_{xy} R
  \ ,
  \\
  \noalign{\medskip}
  R^x{}_{yxy} &=&
  \casehalf g_{yy} R
  \ ,
  \\
  \noalign{\medskip}
  R^y{}_{xxy} &=&
  - \casehalf g_{xx} R
  \ ,
  \\
  \noalign{\medskip}
  R^y{}_{yxy} &=&
  - \casehalf g_{xy} R
  \ .
  \end{array}
  \label{Riemann}
\end{equation}
With $\gamma$ chosen as in the previous section, positive for $\zeta<1$ and
positive imaginary for $\zeta>1$, it is straightforward to
integrate the expressions (\ref{Riemann}) against a test function and to
take the limit $\gamma\to0$. For $\zeta<1$, the resulting nonvanishing
independent components of $R^a{}_{bcd}$ are given by
\begin{mathletters}
\label{riemannlimit}
\begin{eqnarray}
\epsilon=1: &&\ \
R^x{}_{yxy} = - R^y{}_{xxy}
= \pi \left[ { -1 \over (1-\zeta) }
+
{ (2+\zeta) \ln (1-\zeta) \over 2\zeta } \right]
\delta_2(x,y)
\ ,
\label{riemannlimit+}
\\
\epsilon=-1: &&\ \
R^x{}_{yxy} = - R^y{}_{xxy}
= \pi \left[ -1
+
{ (2-\zeta) \ln (1-\zeta) \over 2\zeta } \right]
\delta_2(x,y)
\ ,
\label{riemannlimit-}
\end{eqnarray}
\end{mathletters}%
where at $\zeta=0$ the expressions are to be understood in the sense of
the limit $\zeta\to0$. For $\zeta>1$, the results are obtained from those
in (\ref{riemannlimit}) through replacing $\ln(1-\zeta)$ by
$\ln(\zeta-1)+i\pi$.

Thus we can say that our Riemann tensor, regarded as a distribution acting
on weight one test fields $\Phi_a{}^{bcd}$ that are smooth
in our differentiable structure, has a well-defined
$\gamma\to 0$ limit for each value of the parameters
$\zeta$ and~$\epsilon$.  However, these limits have highly unsatisfactory
properties.  For example, (\ref{riemannlimit+}) does not vanish for
$\zeta=-3$, even though the defect angle then vanishes and the cone
reduces to a plane. Instead, (\ref{riemannlimit+}) vanishes for precisely
two values of~$\zeta$, one corresponding to a positive and the other to a
negative defect angle. Similarly, (\ref{riemannlimit-}) vanishes for
precisely one value of~$\zeta<1$, even though the metric has a defect
angle equal to $-2\pi$ for all values of $\zeta<1$. This highlights
the difficulties discussed in Ref.\ \cite{GT} for defining a distributional
Riemann tensor for conically singular metrics: the distributional limit is
highly sensitive to the choice of the regularization, and specifically to
the choice of differentiable structure that the regularization implicitly
uses.

As a second example, consider the Ricci scalar~$R$. As the Ricci scalar
{\em density\/} for our two-dimensional singular metrics is
proportional to the ordinary delta-function~$\delta_2$, one might attempt
to define the Ricci {\em scalar\/}
$R$ as a distribution proportional to the scalar delta-function
${(\pm g)}^{-1/2} \delta_2$.  The problem with this is that for our metrics
the scalar delta-function is not defined, because the factor
${(\pm g)}^{-1/2}$ is singular.  Note, however, that for the Euclidean
conical singularity of subsection~\ref{subsec:+xismall}, the Ricci scalar
can be defined  as a distribution proportional to the
delta-function if one changes the differentiable structure from that
defined by the coordinate functions
$(x,y)$ to that defined by the functions $(\xi,\eta)$, where
\cite{balasin}
\begin{equation}
\begin{array}{rcl}
  \xi &=& x\sqrt{x^2+y^2}
  \ ,
  \\
  \eta &=& y\sqrt{x^2+y^2}
  \ .
\end{array}
\label{xieta-transf}
\end{equation}
The transformation (\ref{xieta-transf}) is smooth everywhere except at the
origin, where it is only~$C^0$ and its Jacobian diverges.  One can view
this singular Jacobian as canceling the singularity that occurred in the
factor ${(g)}^{-1/2}$ in the coordinates $(x,y)$.\footnote%
{At first sight it might seem paradoxical that the {\em scalar\/} $R$ should
be more sensitive to the choice of differentiable structure than the
{\em density\/}~$\sqrt{g}\,R$.  However, this impression disappears if one
remembers that a distribution is essentially a dual object, and dualization
reverses the roles of scalars and scalar densities.}

\section{Phase of the complex regulator}
\label{sec:imag-sign}

For the yarmulke singularity and the crotch singularity, the sign of the
imaginary part of the Ricci scalar density was fixed by the choice that the
regulator $\gamma$ have a positive imaginary part. In this section
we shall discuss the status of this choice.

In Section \ref{sec:regularized} we deduced the sign of the imaginary part
of $\gamma$ from the condition~(\ref{sharp}), which itself was chosen to
give a positive imaginary part to the action (\ref{Sphi}) of a real
massless scalar field. Including a mass term in the action (\ref{Sphi})
would not have made a difference for $1<\zeta\le2$.
For $\zeta>2$, however, the imaginary parts of the mass term and the
kinetic term in the action have the opposite sign, and the scalar field
path integral on the regularized spacetime is not convergent for either
sign of the regulator.  The same difference between the cases $1<\zeta\le2$
and $\zeta>2$ occurs also for the Maxwell field on a spacetime that is a
product of one of our (1+1)-dimensional metrics with two flat Euclidean
dimensions.

One solution to the problem is to rotate the metric parameter $\zeta$ into
the complex along with~$\gamma$.  In fact, the ansatz $\gamma=i\sigma+\eta$,
$\zeta = \zeta_0-i\eta$, with $\eta=\sqrt{(\zeta_0-2)\sigma}$, $\zeta_0>2$,
takes care of both the electromagnetic and scalar fields, and with it one
still obtains the same limiting values (\ref{Rdens-lor+})
and~(\ref{Rdens-lor2+}) for $\casehalf\sqrt{-g}\,R$ as $\gamma\to0$.  As
far as the scalar field alone is concerned, the difficulty could also be
cured by rotating the coupling constant $m^2$ into the complex, but such a
technique would not help with the Maxwell field, or with nonabelian gauge
fields.

This construction of a regulator may be satisfactory as far as it goes, but,
without further motivation, it seems rather ad hoc, and it doesn't clearly
guarantee that an entirely different choice of regulator might not lead to
a very different outcome.  A more systematic approach would replace
(\ref{sharp}) (or, slightly more generally, the condition
$\Im(g_{ab})\ge0$) by a condition that would guarantee convergence for any
reasonable matter field.  (Linearized gravity is also important, of course,
but is something of a special case \cite{mazur-mottola}.)  It turns out that
if we restrict ourselves to local conditions on small deformations $\delta
g_{ab}$ away from non-degenerate Lorentzian metrics, then an optimum
condition is that the deformation be of the form
\begin{equation}
  \delta g_{ab} = i \sum_\epsilon \eps_a\eps_b
  \ ,
  \label{smallsharp}
\end{equation}
where each covector $\eps_a$ is real and timelike (or null) with respect
to~$g_{ab}$.  This will guarantee (formal) convergence of the path
integral for any field whose classical stress-energy tensor satisfies
(off shell) the weak energy condition ($T^{ab}v_a v_b\ge0, \forall$
timelike $v_a$). Evidently (\ref{smallsharp}) implies (\ref{sharp}), but
is more restrictive.  For finite deformations, there exist analogous
strengthenings of (\ref{sharp}) that work for individual matter fields but
they are harder to state.\footnote%
{It follows from (\ref{sharp}) that one can find a basis of (real) vectors
in which $g_{ab}$ is diagonal,
{\em i.e.,}
in which it assumes the form
$g_{ab}=\sum_j \lambda_j (v_j)_a (v_j)_b$.  The strengthenings referred to
can then be expressed as conditions on the $\lambda_j$.  For the
massless scalar field, we want $\Im(\sqrt{-g}g^{ab})<0$ and hence need
$\Im(\sqrt{- \prod\lambda}/\lambda_j)<0$ for all $j$; with a mass term
present, we need also $\Im\sqrt{-\prod\lambda}<0$.  For the electromagnetic
field the condition is positivity of the imaginary part of the quadratic
form $Q(F):=-(1/4) \sqrt{-g}g^{ab}g^{cd}F_{ac}F_{bd}$, corresponding to the
requirement $\Im(\sqrt{-\prod\lambda}/(\lambda_j\lambda_k))<0$ for all
$j\ne k$.  Here, $g_{ab}$ must be in the same connected component (with
respect to the above conditions) as some real metric of Lorentzian
signature, and the branch of the square root is chosen to be positive for
that metric.  It is easy to verify that all of these conditions are
satisfied along the path of ordinary Wick rotation, which turns
$g_{\mu\nu}={\rm diag}(-1,1,1....1)$ into
$g_{\mu\nu}={\rm{diag}}(1,1,1....1)$ by taking
$g_{00}$ from $-1$ to $+1$ through the upper half plane.  We know of no
general condition analogous to (\ref{smallsharp}) for finite deformations
that is guaranteed to cover all possible matter fields.}
We have not attempted to generalize the infinitesimal condition
(\ref{smallsharp}) or its finite analogues to regions where the metric
degenerates, such as at yarmulke or crotch singularities, but the existence
of the ansatz written down in the previous paragraph strongly suggests this
should be possible.

Even without such a generalization in hand, it seems sufficiently clear
that any adequate regulator must lead to the same values as found in
Section \ref{sec:regularized} for the Ricci scalar density.  Indeed, we
obtained these values by temporarily deforming the metric into the space
``$CI$'' of complex invertible metrics, and the discussion of Section
\ref{sec:gauss-bonnet} shows that the only possible ambiguity associated
with this procedure arises from the choice of which connected component of
$CI$ one deforms into.  But the weakest of the convergence conditions we
have entertained, namely~(\ref{sharp}), already defines a domain
$D\subseteq CI$ that is convex, and therefore connected.  Hence the
Gauss-Bonnet integral must remain constant within~$D$, and no ambiguity can
arise.  (For consistency $\sqrt{-g}$ must be single-valued within~$D$,
which in fact it is, thanks to the convexity of~$D$.)  Clearly, no further
conditions imposed in addition to (\ref{sharp}) can affect this conclusion.
In particular, any regulator respecting (\ref{sharp}) will resolve the sign
ambiguity in (\ref{Rdens-lor}) and (\ref{Rdens-lor2}) in the same way as
led to our basic results, (\ref{Rdens-lor+}) and~(\ref{Rdens-lor2+}).
Essentially the same argument for uniqueness can also be made in the
context of Regge calculus, where the sign of the imaginary part of the
action hinges on how one resolves the complex conjugation ambiguity in the
definition of a Lorentzian angle (cf.\ \cite{sorkin-regge}).  Here again, an
analytic continuation along a path of metrics respecting the condition
(\ref{sharp}) suffices to fix the sign uniquely, and one obtains
contributions to the action entirely consistent with (\ref{Rdens-lor+}) and
(\ref{Rdens-lor2+}), namely $\beta-2\pi{i}$ for the yarmulke and
$\beta+2\pi{i}$ for the trousers.\footnote%
{The second expression refers to a generalized crotch singularity, obtained
from that of Section \ref{sec:regularized}
by removing a wedge of rapidity parameter~$\beta$.
The sign of $\beta$ is positive (respectively negative) if the wedge points
in a spacelike (timelike) direction.}

Further insight into the inevitability of (\ref{Rdens-lor+}) and
(\ref{Rdens-lor2+}) comes from considering their dependence on the
parameter~$\zeta$.  The requirement that a scalar field path integral on
the regularized Lorentzian geometry should be formally convergent is
closely connected with the reasons that mandate the direction of the
Wick rotation in flat-space quantum field theory.  Thus, we might expect
that our results for the crotch and yarmulke singularities could also be
derived from the analogous formulas for conical singularities of Euclidean
signature by viewing the Euclidean and Lorentzian Ricci scalar densities as
analytic continuations of each other in the parameter~$\zeta$.  To see that
this is so, recall that the usual direction of Wick rotation, as specified
for example by~(\ref{sharp}), implies that the Euclidean expression
$\sqrt{g}\,R$ continues to the Lorentzian expression $+i\sqrt{-g}\,R$.
[The relation $iS=-S_{\rm E}$ then leads to the usual definition
\begin{equation}
S_{\rm E} = - {1 \over 2 \kappa} \int d^Dx \sqrt{g}\,R
\ \ + \ \hbox{(boundary terms)}
\label{E-action}
\end{equation}
for the Euclidean counterpart of the Einstein action~(\ref{action}).]
This
same rule that the Euclidean $\sqrt{g}$ continues to the Lorentzian
$i\sqrt{-g}$ is embodied in the formulas given in Section
\ref{sec:regularized} for $\sqrt{g}$ and~$\sqrt{-g}$, as one sees by
setting $\gamma$ to zero, and analytically continuing $\zeta$ past
$\zeta=1$ in the lower half plane (this choice of half-plane being the one
implied by the condition ${\rm Im\,}(g_{ab}) \ge 0$).  Comparing, then,
(\ref{Rdens-dcover}) with $i$ times~(\ref{Rdens-lor2+}), we see that the
crotch curvature density is indeed the analytic continuation of its
counterpart for the Euclidean double cover metric.  The $\epsilon=1$ case
is similar, but slightly more interesting, because of the nontrivial
$\zeta$-dependence of the formulas (\ref{Rdens-e}) and~(\ref{Rdens-lor+}).
Once again we see that $i$ times $\casehalf\sqrt{-g}\,R$ of the yarmulke
and $\casehalf\sqrt{g}\,R$ of the Euclidean general conical singularity
are analytic continuations of each other.  For the signs of the
$\zeta$-dependent terms to agree, $\zeta$ must be continued past $\zeta=1$
in the lower half plane, consistently with what we just observed in
connection with~$\sqrt{g}$.  One can think of the Euclidean defect angle
being continued to the complex value $2\pi+i\beta$, where $\beta>0$ is the
rapidity parameter of the Misner space.  Equivalently, one can say that the
Euclidean opening angle is continued to the purely imaginary
value $-i\beta$.

\section{Conclusions and discussion}
\label{sec:discussion}

In this paper we have investigated two one-parameter families of
$(1+1)$-dimensional topology-changing metrics that contain
Lorentzian analogues of conical singularities.  For the metrics of the
first family, the spacetime is a ``yarmulke" obtained by adding an initial
(or final) singularity to one causal domain of a Misner space.  For the
metrics of the second family, the singularity is that occurring at the
crotch of the Lorentzian ``trousers" spacetime.  Regularizing the metrics
by adding a small positive imaginary part and then taking this
regulator to zero, we found that in both cases the scalar density
$\casehalf \sqrt{-g}\,R$ converges to a delta-function at the
singularity.  For the trousers family the coefficient of the
delta-function is the purely imaginary number~$+2\pi i$, independently of
the parameter; for the yarmulke family it is the complex number $\beta
-2\pi i$, where $\beta$ is the rapidity parameter of the associated
Misner space.

In these coefficients, the signs of the imaginary parts follow from our
having chosen the imaginary part of the regularized metric to be
positive. This property guarantees in particular that a scalar field
functional integral on the regularized spacetime is formally
convergent.

For certain ranges of the parameter $\zeta$ in~(\ref{metric2}), the
metrics of our two families acquire a Euclidean signature, and the
singularity is then of the ordinary conical sort.  In these cases, we
verified that our regularization method, with the regulator chosen real,
renders the scalar density $\casehalf \sqrt{g}\,R$ as a delta-function
concentrated at the singularity, with a strength precisely equal to the
defect angle. (Curiously,
this angle is independent of $\zeta$ for the metrics of the ``trousers''
family.) This is the familiar result that makes the Gauss-Bonnet theorem
hold for Euclidean signature metrics with conical singularities.

In all cases, both Riemannian and Lorentzian, our results are consistent
with the general rule that the contribution to the action integral
(\ref{action}) from a conical-type singularity equals the generalized
defect angle, as defined by the ``complex trigonometry'' natural to Regge
calculus \cite{sorkin-regge}.  This is also what one would obtain from a
suitably complexified Gauss-Bonnet theorem, as discussed in
Section~\ref{sec:gauss-bonnet}. We did not investigate a crotch singularity
with nonzero rapidity parameter~$\beta$, because such a geometry does not
occur among the Morse theory-inspired families of metrics we considered.
However, the results of Sections \ref{sec:regularized} and
\ref{sec:gauss-bonnet} strongly suggest that we would obtain a
delta-function with strength~$\beta+2\pi{i}$ in that case as well.

The main interest of our findings lies in their implications for topology
change. The first issue that arises in this context is whether topological
transitions can proceed as classical processes in  the classical limit of
quantum general relativity.  In $D\ge3$ spacetime dimensions, Einstein
gravity in the metric formulation can be defined in terms of a variational
principle with the action functional~(\ref{action}). Under the usual
smoothness and invertibility assumptions for the metric, stationarity of
the action under appropriate boundary conditions is equivalent to the
vacuum Einstein equations, $G_{ab}=0$.  However, stationarity of the action
$S$ can be regarded, more generally, as a criterion for selecting classical
solutions among all field configurations for which $S$ is defined and
suitably differentiable, even when the metrics contain singularities that
make the interpretation of the Einstein equations as such
problematic.\footnote%
{This reasoning assumes that the field equations of classical gravity
emerge from a quantum action functional in the same way that Maxwell's
equations emerge from the action of the corresponding quantum theory, and
not, for example, in the way the heat equation emerges from the path
integral for the individual molecules.  It is possible that this assumption
is too simple, but if so one would have to explain why the classical theory
has any variational formulation at all.}

In this sense, metrics containing Euclidean conical singularities are known
{\em not\/} to be vacuum solutions to the metric formulation of Einstein
gravity \cite{haylo,cartei1,cartei2,teitel-ext}, the reason being in
essence that the contribution to $S$ from the vicinity of such conical
singularities is proportional to the defect angle in the 2-dimensional
submanifold times the $(D-2)$-dimensional volume of the remaining
dimensions, and this contribution is not stationary under variations of the
$(D-2)$-dimensional volume unless the defect angle vanishes; for details,
see Refs.\ \cite{haylo,cartei1,cartei2,teitel-ext}. The results of this
paper indicate that a similar conclusion holds for spacetimes that contain
Lorentzian singularities of the type we have investigated.  For
example,
the product of one of our singular 2-metrics with a flat torus of
$D-2$ dimensions cannot be regarded as a solution of Einstein gravity in
the metric formulation.\footnote%
{Specifically, the action $S$ of such a product is not stationary under
certain variations that rescale the metric of the torus by a scaling
function of compact support.  In fact, if we choose the scaling function to
be constant in a neighborhood of the singularity, then the spacetime
retains its product form in that neighborhood (globally it is a ``warped
product'') and the Ricci scalar density remains a delta-function there,
making it easy to compute the overall action.  No doubt there are many more
(and more localized) variations under which $\delta S \not= 0$, but to
render such a statement meaningful one would first have to extend the
analysis performed in this paper to define $S$ for the more general class
of conically-singular metrics to which such variations would
lead. \hfil\break
\hbox to 10 pt{\hfill}At a very technical level, there arises in this
connection the issue of how even to define a differentiable structure for
the enlarged class of metrics under consideration. (This issue would become
entangled with that of asymptotic boundary conditions, were one to allow
the conical type singularities to extend out to infinity, a point
emphasized to us by Abhay Ashtekar.)  However, it may be premature to try
to resolve such technical issues at this point.  Does one even expect the
relevant space of generalized metrics to form a smooth manifold for
example?  The criterion for calling a geometry a classical solution should,
in our view, ultimately be chosen to express the condition that the paths
in its neighborhood make a non-negligible contribution to the gravitational
functional integral, but the precise meaning of this criterion, not to
mention its precise relation to attributes such as differentiability of the
classical action, is far from settled at this point, all the more so if a
fundamental discreteness proves to be required before the gravitational
functional integral can be given precise meaning. For a discussion of
issues related to the above, see Refs.\
\cite{haylo,LW1,cartei1,cartei2,teitel-ext,ash-vara,LW2}. }

If topological transitions thus are forbidden in the classical limit, the
next question that arises is whether they can proceed as quantum tunneling
processes.  Let us consider this question from the vantage point of
path-integral quantization.  As our product metrics are not stationary
points of the action~$S$ in $D\ge3$ dimensions, one would not expect the
path integral to gain an appreciable contribution from them.  However, the
situation is complicated by the fact that $S$ is now complex.  On one hand
this means that destructive interference will be associated only with
non-stationarity of the {\em real part\/} of~$S$.  (Non-stationarity of
$\Im(S)$ entails no suppression as such, although it does imply the
presence of nearby paths with amplitudes of greater absolute magnitude.)
On the other hand, the very existence of an imaginary part can now lead to
its own suppression or enhancement.  Indeed the action (\ref{action}) has
an imaginary part proportional to the $(D-2)$-dimensional volume of the
singularity, and this
$(D-2)$-dimensional volume may be large.  For the crotch singularity the
imaginary part of the action is positive, and the contribution to the path
integral from the trousers spacetimes is therefore {\em suppressed\/} by an
exponential factor.\footnote%
{There is an obvious, but potentially misleading, analogy here with
tunneling solutions, whose complex action also entails suppression.  In
that case, however, the history or path is itself complex and therefore
without direct physical meaning, being of interest only as a saddle point
of the analytically extended amplitude.  In contrast, the paths under
consideration in this paper represent possible histories of the actual
gravitational field, even if they are not extrema of~$S$.}
Topology change via the trousers mechanism is therefore very strongly
disfavored, the suppression factor in $3+1$ dimensions being~$\exp(-A/4G)$,
where $A$ is the area of the two flat dimensions.  For the yarmulke
singularity, in contrast, the imaginary part of the action is negative, and
the contribution to the path integral from the yarmulke spacetimes is
therefore {\em enhanced\/} by an exponential factor.  One may perhaps take
this as evidence for an exponential enhancement of the creation (or
annihilation) of a universe by such a mechanism.\footnote%
{Is there any useful analogy between this exponentially large factor and the
similar factor in the wave function that Hartle and Hawking obtained from
the no-boundary path integral in the positive curvature Friedmann model
with a cosmological constant \cite{vatican,HH,JJHlouko1}?
Likewise, is there more than a formal significance in the fact that the
coefficient of the area that enters here ($8\pi/\kappa=1/4G$) is precisely
the same one that occurs in computing the entropy of a black hole horizon
(cf.\ \cite{cartei1,cartei2,SSS})?} However, one would need to balance this
enhancement against the suppression coming (when $D\ge 3$) from the fact
that (unlike for the trousers) the yarmulke spacetimes have a Ricci
scalar density with non-zero real part, and therefore are not stationary
points of~$\Re(S)$.

A second implication of our results concerns two dimensional spacetimes as
such, where the effect of the action being complex is still present,
even if it is not exponentially large.  In two dimensions,
the relative enhancements and suppressions attaching to the
yarmulke and trousers are strongly reminiscent of the topological expansion
in string theory.  However, in that situation there seems to be no strong
evidence about the sign of the coupling constant~$\kappa$, unlike in
(3+1)-gravity, where not only stability, but experiment dictates that
$\kappa$ be positive.  In any case, it seems worthwhile to ask whether the
findings of this paper have any significance for strings.

Our results for the metric formulation provide a striking contrast to the
vielbein-cum-connection formulation of Einstein gravity, where a spacetime
obtained as the product of one or more flat dimensions with (for example)
the double cover conical singularity or the crotch singularity can be
regarded as a solution to the classical field equations \cite{horo}.
Formally, this difference stems from the different roles played on
the one hand by the Christoffel connection in the metric formulation, and
on the other hand by the Lorentz connection in the vielbein-cum-connection
formulation.  Our metric action for singular metrics was defined in terms
of a limiting process through nonsingular metrics.  Since the Christoffel
connection is uniquely determined by the metric regardless of whether the
field equations hold, the (possibly singular) limiting behavior of the
Christoffel connection upon approaching our singular metrics is uniquely
determined by the limiting process.  In the connection formulation, on the
other hand, the relation between the Lorentz connection and the vielbein
is an equation of motion, and this equation is well defined
even when the vielbein is degenerate.  One can therefore directly look for
solutions to the full field equations containing singular vielbeins,
without having to interpret the latter by reference to a regularized
vielbein field.  We shall illustrate this in Appendix \ref{app:dcon} for
a $(2+1)$-dimensional spacetime that is the product of the Euclidean
double cover conical singularity with a time dimension \cite{horo}, and
in Appendix \ref{app:hourglass} for the Lorentzian ``hourglass"
spacetime \cite{horoPLB,horoMG6}.  However, this formal analysis of why
the two formulations differ leaves open the physical meaning of the
difference. In the metric formulation the trousers does not provide a
classical route to topology change, whereas in the
vielbein-cum-connection formulation, it apparently does.  How can these
conclusions be reconciled?

There would seem to be at least three possible answers to this question.
First, it might be that, properly understood, the conditions for a
configuration to represent a possible classical evolution are not actually
satisfied by the putative solutions, even in the vielbein-cum-connection
version of Einstein gravity.  This might occur, for example, through some
subtle failure of the putative solution to possess a neighborhood of
sufficient functional measure, but we have not been able to find any
convincing reason why that should be the case.
(However,  the situation appears to be subtle;
compare the discussion of ``$C^0$ isometries'' below.)
A~second possibility is that the
trousers {\em is\/} a solution, but that it belongs to an entirely
disjoint sector of history space, such that the vielbein-cum-connection
formulation agrees with the metric formulation when these sectors are
excluded by hand. The third possibility is that the two formulations are
irredeemably different, with one allowing classical topology change and
the other forbidding it. In either of the last two cases, one must decide
which formulation is physically correct, and here it seems the
experimental evidence strongly favors the metric formulation.  Otherwise,
we would have no apparent explanation for why the {\em macroscopic\/}
topology of spacetime is not incessantly and prolifically changing in a
radically unpredictable manner \cite{horo}. (Assuming such changes could
proceed with anything like a classical amplitude, the shear variety of
possibilities would seem to make their occurrence inevitable on entropic
grounds.)  If borne out, this conclusion would present one of the rare
instances where one can experimentally decide between contending theories
of quantum gravity.

Taken as a whole, the findings reported here provide convincing, if
indirect, evidence in favor of our complex expressions for the integrated
 scalar curvature.  What is lacking
is a deeper understanding of the
origin of the imaginary part of these integrals, and more generally of the
significance of the complex regularizations we have employed.  To
provide a straightforward meaning to our regularized action in the
context of the Lorentzian sum-over-histories, one would have to first
define a purely Lorentzian path integral incorporating non-invertible
metrics of the type we have been considering, and then show that a
conical-type singularity in the metric introduces an extra suppression or
enhancement factor, either because the amplitude itself is not of unit
modulus or (to the extent that the distinction makes sense) because the
measure factor is changed by the presence of the singularity.  We suspect
that such a demonstration is not fully feasible within a continuum
theory, because the path integral itself is ultimately not definable
there, but that something of the sort could emerge from a theory with a
built in discreteness, such as provided by a causal set account of
spacetime.

In this connection, it seems natural to look for some relationship between
our complex action and the divergent stress-energy tensor induced by a
crotch or yarmulke singularity via its influence on the quantum matter
fields (or gravitons) in its neighborhood
\cite{raf-vict,anderson-dewitt,harris-dray,dray+,raf-fluct,horo}.  The
suppression of topology change one would infer from these radiative
effects seems reminiscent of the exponential suppression discussed above
coming from Im($S$) (but what of the enhancement in the yarmulke case?).
Similarly, such radiative effects might help resolve the discrepancy
discussed earlier between the metric formulation of quantum gravity and
the vielbein-cum-connection formulation.

Finally, we would like to raise two general questions that are suggested by
the work described above.

The first question is one of principle touching the meaning of ``general
covariance'' in the context of non-invertible metrics such as we have been
concerned with.  Within the four families of spacetimes we have been
working with, there are many examples of pairs of metrics that represent
the ``same'' geometry (and share the same action and distributional scalar
curvature density), but nevertheless are not isometric via any
diffeomorphism.   For example, the metric of subsection
\ref{subsec:+xismall} becomes a cone with zero defect angle when
$\zeta=-3$, but, as discussed in Section~\ref{sec:curvature}, there is no
{\em smooth\/} and smoothly invertible transformation of the coordinates
$x$ and $y$ that would bring this metric to an explicitly flat Euclidean
form (cf.\ equation~(\ref{transf1})).
Similarly, all of the metrics of the ``trousers'' family of subsection
\ref{subsec:-xilarge} are characterized by a conical singularity of the
same strength, yet again there exists no diffeomorphism taking one value
of $\zeta$ to another.  In the context of invertible metrics, we are used
to defining a ``geometry'' as a diffeomorphism equivalence class of
metrics, and declaring that any two metrics that define the same
geometry should be identified physically (``general covariance'').
But if we want to make a physical identification among all of our trousers
metrics, or among all of our double cover conical singularity metrics, then
a broader notion of equivalence must be adopted.  To do otherwise would
lead to a multiple counting of such ``histories'' in the path integral,
which presumably would be incorrect.

Thus we arrive at the view that the class of valid ``gauge
transformations'' is not exhausted by the diffeomorphisms.  Rather a more
general subset (seemingly not a sub{\it group}!) of the homeomorphism group
seems to be indicated, resulting in a more general notion of equivalence
that might be termed ``generalized (or $C^0$) isometry''.  Said another
way, this means that the same geometry can be given more than one
differentiable structure, without any genuine physical change having
occurred. This may be seen as suggesting that the differentiable
structure has in some ways less ``reality'', than, say, the topology or
the metric itself.  But how, precisely, should one define the concept of
generalized isometry?  In a Riemannian setting, there is apparently no
trouble, since one can just define a generalized isometry to be any
homeomorphism preserving the global distance function, which remains
well defined even in the presence of singular points where the metric
degenerates or vanishes.  In the Lorentzian setting, on the other hand,
it is not so obvious how to proceed.  Plausibly, the appropriate
definition is that a generalized isometry is a homeomorphism
$f$ that preserves both the local causal order and the volume element
$\sqrt{-g}$ (cf.\ \cite{luca-raf}). These requirements are meaningful for
the type of metric we studied in Section~\ref{sec:regularized}, and they
offer a natural generalization of the usual definition, since they
can be shown to imply that $f$ is an isometry in the ordinary sense when
both metrics involved are smooth and invertible.

Finally, we may ask to what extent the complex actions we have found are
characteristic features of topology-changing spacetimes in general, and to
what extent they are peculiar to two dimensions and the particular family
of metrics we have chosen to study.  Concerning dimensionality, we may
observe that the scalar curvature of metrics like those of Section
\ref{sec:regularized} diverges no more strongly in higher dimensions than
it does in two.  But this means that the Ricci scalar {\em density\/} is
less singular, so that one might anticipate less of a need for any sort of
regularization in order that the Hilbert action $S$ be defined.  We might
thus expect that the occurrence of an imaginary part of this $S$ is indeed
peculiar to (1+1)-dimensional spacetimes.  On the other hand, we may also
anticipate that within two dimensions, a complex value for $S$ is an
unavoidable concomitant of topology change.  This expectation derives from
the observation that topology change entails a modification of the light
cone structure, and this change can be interpreted in terms of a complex
defect angle in the sense discussed in Section~\ref{sec:gauss-bonnet}.  And
because of the generality of the considerations of that section, one would
expect a corresponding generality in the conclusion that a complex defect
angle implies a complex action.
(Similarly, we would expect the Lovelock actions
\cite{lovelock1,lovelock2} to acquire well-defined imaginary parts
as a result of topological transitions in higher dimensions.)

\acknowledgments
We would like to thank Arley Anderson, Abhay Ashtekar, Luca Bombelli,
Arvind Borde, Tevian Dray, John Friedman, Dmitri Fursaev, Domenico
Giulini, Gary Horowitz, Ted Jacobson, Klaus Kirsten,  Jean Krisch, Don
Marolf, and Don Page for useful discussions and correspondence. We would
also like to thank Paul Ehrlich, Steve Harris, and Adam Helfer for
bringing Refs.\
\cite{jee,birman,law} to our attention.
This work was supported in part by
the NSF grants
PHY-90-16733,
PHY-91-05935,
PHY-93-07570,
PHY-94-21849,
and
PHY-95-07740,
and by research funds provided by Syracuse University.

\appendix

\section{Double cover conical singularity
in the triad-cum-connection formulation of (2+1)--dimensional gravity}
\label{app:dcon}

In this appendix we shall consider, in the triad-cum-connection
formulation of Einstein gravity (sometimes also called the ``first order
formalism'' or ``Cartan formulation''), the $(2+1)$-dimensional spacetime
that is obtained as the product of the time axis with the double cover
conical singularity of subsection~\ref{subsec:-xismall}.  We first review
how this spacetime can be regarded as an exact solution in this
formulation.  We then demonstrate that if one attempts to approach this
spacetime via a specific class of nondegenerate triads, such that the
equation of motion relating the connection to the triad is imposed, the
connection does not have a limiting value, and its curvature approaches a
delta-function rather than zero. Moreover, even neglecting the
connection, the sequence of non-degenerate triad fields does not go over
as $\gamma\to0$ to the triad field of the solution.

Recall that in the triad-cum-connection formulation of $(2+1)$-dimensional
gravity of Refs.\ \cite{witten1,AAbook2,romano,amano}, the fundamental
variables are the co-triad $e_{aI}$ and an $\so$ connection~$A^I_a$.
We follow the notation of Refs.\ \cite{AAbook2,romano}
(for these appendices only).
Assuming the
spacetime to be orientable, the action is
\begin{equation}
S = {1 \over 2\kappa} \int {d^3x} \,
\tilde{\eta}^{abc}
\, {e}_{aI} {F}^I_{bc}
\ \ ,
\label{3-action}
\end{equation}
where $\tilde{\eta}^{abc}$ is the Levi-Civita density and
${F}^I_{ab}$ is the curvature of the connection,
\begin{equation}
{F}^I_{ab} =
2 \partial^{\phantom{I}}_{[a} {A}^I_{b]} +
\epsilon^I{}_{JK} {A}^J_a {A}^K_b
\ .
\label{curvform}
\end{equation}
The field equations are
\begin{mathletters}
\label{eom3d}
\begin{eqnarray}
&&F^I_{ab} =0
\ ,
\\
&& {\cal D}_{[a} {e}_{b]I} = 0
\ ,
\label{3-eom2}
\end{eqnarray}
\end{mathletters}%
where ${\cal D}_a$ is the gauge covariant derivative determined by
${A}^I_a$,
\begin{equation}
{{\cal D}}_a v_K =
\partial_a v_K -
\epsilon^I{}_{JK} {A}^J_a v_I
\ \ .
\end{equation}

If the triad is invertible, equations (\ref{3-eom2}) can be uniquely
solved for the connection.  Inserting this solution back into the
action (\ref{3-action}) yields the action of the usual metric theory
expressed in terms of the triad,
\begin{equation}
S = {1 \over 2\kappa} \int {d^3x}
\sqrt{-g} \, R
\ .
\label{3-action-metric}
\end{equation}

We wish to consider the spacetime metric
\begin{equation}
ds^2 = - dt^2 +
\left( x^2 + y^2 + \gamma \right)
\left( dx^2 + dy^2 \right)
\ ,
\label{3-metric}
\end{equation}
where $\gamma\ge0$.  For $\gamma=0$ this metric is just the product of
the time axis and the double cover conical singularity metric
studied in subsection~\ref{subsec:-xismall}.  It is well known that the
conical singularity metric can be regarded as a solution to the
triad-cum-connection formulation \cite{horo}; an explicit connection and
triad are given by $A^I=0$ and
\begin{equation}
\begin{array}{rcl}
e^0 &=& dt
\ ,
\\
e^1 &=& xdx - ydy
\ ,
\\
e^2 &=& xdy + ydx
\ .
\end{array}
\label{triad1}
\end{equation}

Let then $\gamma>0$.  We choose a triad compatible with
(\ref{3-metric}) to be
\begin{equation}
\begin{array}{rcl}
e^0 &=& dt
\ ,
\\
\noalign{\smallskip}
e^1 &=& \sqrt{x^2 + y^2 + \gamma} \, dx
\ ,
\\
\noalign{\smallskip}
e^2 &=& \sqrt{x^2 + y^2 + \gamma} \, dy
\ .
\end{array}
\label{triadr1}
\end{equation}
Solving (\ref{3-eom2}) for the connection yields
\begin{equation}
A^0 = {xdy -ydx \over x^2 + y^2 + \gamma}
\ ,
\end{equation}
and the curvature form (\ref{curvform}) is then given by
\begin{equation}
F_{xy}^0 = {2\gamma \over {(x^2 + y^2 + \gamma)}^2}
\ .
\end{equation}
In the limit $\gamma\to0$, $F^0_{xy}$ tends to $2\pi \delta_2(x,y)$,
as can be verified by a straightforward integration against a test
function.  Thus, in this limit one does not recover a solution to the
connection formulation, but instead one obtains a singular connection
with a nonvanishing distributional curvature.  Comparing
(\ref{3-action}) and (\ref{3-action-metric}) shows that this
distributional curvature agrees
with the result (\ref{Rdens-dcover}) for the Ricci scalar density
in the metric formulation.

The failure of the $\gamma\to0$ limit to yield a solution to the
connection formulation
is related to the difference
between (\ref{triad1}) and the $\gamma\to0$ limit of~(\ref{triadr1}),
which are
connected
by a singular gauge transformation that changes the ``winding number'' of
the triad around closed loops that encircle the $t$-axis.  Indeed this
difference of winding number seems to imply that the solution
configuration cannot be obtained as the limit of any sequence of
non-degenerate configurations.  In that sense it would belong to an
entirely different component of ``history space'' than the regular
configurations.

\section{The hourglass spacetime}
\label{app:hourglass}

In this appendix we consider the $(2+1)$-dimensional metric
\begin{equation}
ds^2 = - dt^2
+ \left(t^2 + \gamma \right) d\varphi^2
+ dz^2
\ ,
\label{hour-metric}
\end{equation}
where $t$ and $z$ take all real values, $\varphi$ is periodic with
period $\beta>0$, and $\gamma\ge0$. For $\gamma>0$ the metric is
invertible everywhere. For $\gamma=0$, one obtains the hourglass
spacetime \cite{horoPLB,horoMG6}: the regions $t<0$ and $t>0$ of the
constant $z$ slices are isometric to  past and future causal
regions of Misner space \cite{haw-ell}, but at $t=0$ the metric is
degenerate.

For $\gamma>0$, the scalar density $\casehalf \sqrt{-g}\,R$ in the
chart $(t,\varphi,z)$
is given by
\begin{equation}
\casehalf \sqrt{-g}\,R =
{\partial^2 \over {\partial t}^2}
\left( \sqrt{t^2 + \gamma}\right)
\ ,
\end{equation}
which has at $\gamma\to0$ the distributional limit
\begin{equation}
\casehalf \sqrt{-g}\,R = 2 \delta(t)
\ .
\label{hour-Rdens}
\end{equation}
Integrating (\ref{hour-Rdens}) over the whole spacetime
yields~$2\beta$, which is twice the real part of the corresponding
integral for the yarmulke spacetime considered
in subsection~\ref{subsec:+xilarge}. {}From this point of view, one can
envisage the hourglass spacetime as consisting of two
copies of the yarmulke spacetime, with one flat
dimension added: the yarmulkes have been glued together back to back at the
singularity, in such a way that the real parts of the actions add and the
imaginary parts cancel.

Consider now the triad-cum-connection description. For $\gamma=0$, the
metric (\ref{hour-metric}) can be regarded as a
solution \cite{horoPLB,horoMG6}: a~compatible triad and connection are
given by
\begin{mathletters}
\begin{equation}
\begin{array}{rcl}
e^0 &=& dt
\ ,
\\
\noalign{\smallskip}
e^1 &=& t d\varphi
\ ,
\\
\noalign{\smallskip}
e^2 &=& dz
\ ,
\end{array}
\label{hour-triad1}
\end{equation}
and
\begin{equation}
A^2 = d\varphi
\ .
\end{equation}
\end{mathletters}%

Suppose then that $\gamma>0$, and consider the triad
\begin{equation}
\begin{array}{rcl}
e^0 &=& dt
\ ,
\\
\noalign{\smallskip}
e^1 &=&
\sqrt{t^2 + \gamma} \, d\varphi
\ ,
\\
\noalign{\smallskip}
e^2 &=& dz
\ ,
\end{array}
\label{hour-triadr1}
\end{equation}
which clearly gives the metric~(\ref{hour-metric}).
When (\ref{3-eom2}) holds,
the connection compatible with (\ref{hour-triadr1})
is given by
\begin{equation}
A^2_\varphi =
{\partial \over {\partial t}}
\left( \sqrt{t^2 + \gamma}\right)
\ ,
\label{hour-conn1}
\end{equation}
and the curvature form by
\begin{equation}
F^2_{t\varphi} =
{\partial^2 \over {\partial t}^2}
\left( \sqrt{t^2 + \gamma}\right)
\ .
\end{equation}
In the limit $\gamma\to0$,
we obtain the non-differentiable triad
\begin{mathletters}
\label{hour-fields2}
\begin{equation}
\begin{array}{rcl}
e^0 &=& dt
\ ,
\\
\noalign{\smallskip}
e^1 &=& |t| d\varphi
\ ,
\\
\noalign{\smallskip}
e^2 &=& dz
\ ,
\end{array}
\label{hour-triad2}
\end{equation}
and the discontinuous connection
\begin{equation}
A^2 = {\rm sgn}(t) d\varphi
\ .
\end{equation}
\end{mathletters}%
The curvature has the nonvanishing distributional limit
\begin{equation}
F^2_{t\varphi} = 2\delta(t)
\ \ .
\label{F-dist}
\end{equation}
Equation (\ref{F-dist}) can also be inferred from the non-smooth
fields (\ref{hour-fields2}) directly, without going through the
regularization \cite{horo-private}.\footnote%
{We thank Gary Horowitz
for this observation, which first brought the hourglass spacetime
to our attention.}
Thus, although the fields (\ref{hour-fields2}) are compatible with
the hourglass spacetime, they cannot be regarded as a solution to the
connection formulation.
Indeed they seem to belong, as before, to an
entirely different component of the space of field configurations.
Again, comparing (\ref{3-action}) and
(\ref{3-action-metric}) shows that the result (\ref{F-dist})
is in agreement with the result (\ref{hour-Rdens}) in the metric
formulation.

\newpage


\end{document}